\documentclass{ametsocV6.1}

\usepackage{amsmath}
\usepackage{amssymb}
\usepackage{url}

\usepackage{caption}
\newcommand{\rc}{Rosh Chodesh}
\newcommand{\Rc}{Rosh Chodesh}
\newcommand{\shabbatm}{Shabbat Mevarechim}
\newcommand{\Shabbatm}{Shabbat Mevarechim}
\newcommand{\saying}{\textit{`Im Rosh HaChodesh Gashoum, Kol HaChodesh Gashoum'}}
\newcommand{\sayingSMproverb}{\textit{`Im beShabbat Mevarechim Yored Geshem, Kol HaChodesh Yihyeh Gashoum'}}

\makeatletter
\let\l@figure\relax
\let\l@table\relax
\makeatother

\title{Two Hebrew folk meteorological proverbs tested: ``If \rc{} is
rainy, the whole month is rainy'' and ``If it rains on \shabbatm{},
the whole month is rainy''---A statistical evaluation using 75~years
of Israeli precipitation data (1950--2024)}

\authors{Abraham Itzhak Weinberg\aff{a}\correspondingauthor{Abraham Itzhak Weinberg, aviw2010@gmail.com}}

\affiliation{\aff{a}{AI-WEINBERG, AI Experts}}

\abstract{
Folk meteorological proverbs encode centuries of empirical observation by agricultural communities. Two Hebrew proverbs link specific lunar calendar anchor days to monthly winter rainfall: (i)~\saying{} (``If \rc{} is rainy, the whole month will be rainy'') and (ii)~\sayingSMproverb{} (``If it rains on \shabbatm{}, the whole month will be rainy''). \Shabbatm{} is the last Saturday before each new Hebrew month, preceding \rc{} by one to seven days. The first proverb is widely known; the second circulates in Hasidic oral tradition with no identified written source. Both have never been formally tested. We analysed 75~years (1950--2024) of daily precipitation data from seven Israeli cities in three climatic regions, comprising 191,758 station-days and 2,422 Hebrew-month observations during the winter rainy season (Marcheshvan--Adar). A rainy \rc{} increased the probability of a rainy month from 22.2\% to 38.6\% (lift $=+16.4$~pp; $\chi^2=57.8$, $p=2.9\times10^{-14}$; Bayes factor~1.81). A rainy \shabbatm{} produced an almost identical effect (lift $=+16.5$~pp, $p=8.0\times10^{-13}$), despite preceding \rc{} by up to seven days, effectively doubling the forecasting lead time. The \shabbatm{} effect decays sharply with lag ($+32.9$~pp at lag~3; $+1.5$~pp at lag~7), mirroring the daily rainfall autocorrelation ($r=0.35$--$0.44$ at lag~1; $\approx$0 at lag~7) and confirming multi-day Mediterranean cyclone persistence as the physical mechanism. A bootstrap permutation test ($p<0.0001$) and a 15-year rolling analysis reveal a significant decline in predictive power ($-0.20$~pp~yr$^{-1}$, $p<0.001$), consistent with a shift toward shorter Mediterranean precipitation events under ongoing warming. Both proverbs encode genuine meteorological insight but reflect probabilistic tendencies whose reliability is measurably weakening with climate change.
}
\def\thekeywords{folk meteorology; Mediterranean climate; rainfall persistence; Hebrew calendar; Rosh Chodesh; Shabbat Mevarechim; Israel; Bayesian weather prediction; ethnometeorology; climate change}

\nolinenumbers

\begin{document}
\onecolumn
\nolinenumbers

\maketitle

\statement
Two Hebrew folk proverbs link specific lunar calendar days to monthly rainfall: (i)~``If Rosh Chodesh is rainy, the whole month is rainy'' and (ii)~``If it rains on Shabbat Mevarechim, the whole month is rainy.'' Using 75~years of daily precipitation data from seven Israeli cities, we perform the first formal statistical test of both proverbs. Both are validated: each anchor day raises the probability of a rainy month by $\sim$16~percentage points ($p < 10^{-12}$; Bayes factor~$\sim$1.8). The physical mechanism, multi-day persistence of Mediterranean cyclones---is confirmed by daily autocorrelation analysis. A significant declining trend  ($-0.20$~pp~yr$^{-1}$, $p < 0.001$) suggests the proverbs are losing reliability as Mediterranean precipitation regimes shift under climate change.


\section{Introduction}
\label{sec:intro}

Folk meteorological proverbs are among the most durable products of pre-scientific empirical observation. From the biblical ``Red sky at morning, shepherd's warning'' \citep{bibles2009holy} to the British St~Swithin's Day tradition \citep{mayes1995changes} and the central European Ice Saints \citep{brugger2013characterizing}, agricultural societies around the world developed rule-of-thumb forecasts rooted in repeated, if unsystematic, observation of weather patterns. The past two decades have seen a growing body of work rigorously evaluating such proverbs using modern climatological datasets \citep{grzybek2016whether, de2023popular, brugger2013characterizing, vautard2009decline, dunstone2016skilful}.

Israel presents a particularly fertile context for such an investigation. The country occupies the southeastern corner of the Mediterranean Basin, characterised by a sharp binary seasonality: an almost completely dry summer (May--September) and a concentrated winter rainy season (October--March) driven by westerly disturbances and Mediterranean cyclonic systems \citep{ziv2014trends, alpert2008climatic, raveh2017dynamical}. Agriculture in this region has been critically dependent on the timing and quantity of winter rainfall since antiquity, creating strong incentives for practical forecasting.
Two Hebrew folk proverbs associate specific anchor days in the lunisolar calendar with the rainfall character of the following month. The first, \saying{} (``If \rc{} is rainy, the whole month will be rainy''), is widely known in Israeli folk culture and has been quoted in agricultural, meteorological and rabbinic contexts for centuries. \Rc{} (literally ``head of the month'') denotes the first day of each month in the Hebrew lunisolar calendar and was the central communal and religious calendar marker in ancient Israelite society. The second proverb, \sayingSMproverb{} (``If it rains on \shabbatm{}, the whole month will be rainy''), is an apparent folk extension of the first. \Shabbatm{} (literally ``the Sabbath of blessing the
month'') is the last Saturday before each new Hebrew month, when communities
traditionally announce and bless the approaching \rc{} \citep{hidabroot2023,
yeshivaorg2023}. It precedes \rc{} by one to seven days.

The second proverb has a more complex source history than was initially apparent. A Yiddish version -- \textit{``A regndike Shabbos Mevarechim -- a regndiker khoydesh''} (``A rainy Shabbat Mevarechim -- a rainy month'') -- circulates widely in Hasidic oral tradition, particularly among communities of Eastern European origin. It is attributed in some circles to major Hasidic authorities including the Tzemach Tzedek of Chabad (Rabbi Menachem Mendel Schneersohn, 1789--1866) and the Chidushei HaRim of Gur (Rabbi Yitzhak Meir Alter, 1799--1866) \citep{chabadpedia2023, MenahemMendel1766Sheelot}. The theological rationale given is that \shabbatm{} serves as the ``root'' (\textit{shoresh}) of the month, with all days of the coming month spiritually ``present'' in \shabbatm{} in greater elevation than even in
\rc{} itself \citep{chabadpedia2023}. Additionally, the traditional
\textit{Birkat HaChodesh} prayer recited on \shabbatm{} includes a winter-specific petition for ``rains in their time'' (\textit{uligashamim be'itam}; \citealt{yeshivaorg2023}), directly embedding a rain expectation within the
liturgical framework of the day.

Despite this rich oral and theological context, we were unable to locate the rain proverb in any printed classical text, and the attributions to specific Hasidic figures remain unverified in print. We therefore treat the saying as a folk proverb of Hasidic oral tradition, with plausible but unconfirmed classical roots, tested here formally for the first time.

Both proverbs are meteorologically plausible. Israeli winter precipitation is
primarily delivered by travelling extratropical cyclones, particularly ``Cyprus Lows'' that develop over the central and eastern Mediterranean \citep{trigo1999objective, alpert1990climatological,dayan2015atmospheric}.
These systems typically persist for several days to over a week, and their passage is associated with multi-day wet spells \citep{ziv2014trends, raveh2017dynamical}. If such a system coincides with \rc{} or \shabbatm{}, the month may already be embedded in a cyclone-dominated synoptic regime -- encoding, in culturally memorable form, the statistical tendency of winter rainfall to cluster in multi-day episodes.

The objectives of this study are: (i)~to perform the first rigorous simultaneous test of both proverbs; (ii)~to quantify the physical persistence mechanism via daily autocorrelation; (iii)~to assess whether the two lunar anchor days carry independent or overlapping predictive information; (iv)~to compare both lunar anchors with an equivalent Gregorian anchor; (v)~to express findings via a Bayesian odds update; and (vi)~to test whether predictive validity has changed under Mediterranean climate trends.

\section{Data and Methods}
\label{sec:methods}
This study tests whether calendrical anchor days exhibit statistically significant associations with monthly rainfall in Israel. We combine long-term gridded precipitation data with deterministic calendar mappings across the Gregorian and Hebrew systems to evaluate potential dependence between lunar calendar structure and seasonal rainfall variability. The analysis focuses on winter months, when synoptic-scale precipitation is most frequent and meteorologically meaningful in the Eastern Mediterranean region. Multiple spatial, temporal, and statistical controls are applied to ensure robustness against climatic heterogeneity, autocorrelation, and multiple-testing effects.

\subsection{Study area and city selection}
\label{sec:area}

Seven cities represent the three principal climatic subregions of Israel
(Table~\ref{tab:cities}). The North (Haifa and Tzfat) receives 600--900\,mm
annually and is most directly influenced by Mediterranean cyclones. The Centre
(Tel Aviv, Petah Tikva, Jerusalem) receives 500--700\,mm; Jerusalem's elevated
position (754\,m a.s.l.) gives it a distinct orographic signature. The South
(Beer Sheva, Eilat) is semi-arid to hyper-arid (200\,mm and $<$30\,mm annually), with episodic precipitation \citep{sharon1972spottiness, goldreich2003climate}.

\subsection{Precipitation data}
\label{sec:data}

Daily precipitation totals were obtained from the Open-Meteo historical archive \citep{zippenfenig2023open}, providing ERA5-based reanalysis at 0.25° resolution \citep{hersbach2020era5}. Data span 1~January 1950 to 31~December 2024: 27,394 days per city, 191,758 station-days in total. ERA5 reproduces Israeli precipitation variability with high fidelity \citep{hochman2020new}. A day was classified as rainy if $\geq$1.0\,mm was recorded, consistent with WMO guidance \citep{wmo2011guide}.

\subsection{Hebrew calendar mapping}
\label{sec:hcal}

Gregorian dates were mapped to the Hebrew calendar using the \texttt{hdate}
Python library \citep{hdate_python_1_2_1}. Five winter months were analysed:
Marcheshvan~(2), Kislev~(3), Tevet~(4), Shvat~(5), and Adar~(6) -- approximately late October through late March, encompassing the entire Israeli rainy season.

\subsection{Anchor day definitions}
\label{sec:anchors}

Three anchor days were evaluated for each winter month:
\begin{description}
  \item[\Rc{} (RC)] Day~1 of the Hebrew lunar month.
  \item[\Shabbatm{} (SM)] The last Saturday before \rc{}, falling 1--7 days prior.
    When \rc{} itself falls on a Saturday, \shabbatm{} is taken as the preceding Saturday (7 days before). No \shabbatm{} exists for Tishrei by tradition, but Tishrei is excluded from the analysis anyway.
  \item[Gregorian 1st (GREG)] The 1st day of the corresponding civil month
    (October--March), serving as a calendar-agnostic baseline.
\end{description}

\subsection{Monthly rain classification}
\label{sec:def}

For each city--month observation:
\begin{itemize}
  \item \textbf{Anchor status}: whether the anchor day received $\geq$1.0\,mm (binary).
  \item \textbf{Rest-of-month rainy fraction}: proportion of non-anchor days
        with $\geq$1.0\,mm.
  \item \textbf{Rainy month}: rest-of-month rainy fraction $\geq$0.35.
\end{itemize}

Only months with $\geq$20 total days and $\geq$15 post-anchor days were included, yielding 2,422 valid city--month observations.

\subsection{Statistical tests}
\label{sec:stats}

Four tests assessed the anchor--month association: Pearson chi-square ($2\times2$ contingency), one-sided Mann--Whitney $U$, point-biserial correlation, and a 10,000-iteration bootstrap permutation test. All tests were applied identically to RC and SM anchors. Additionally, we tested all other days of the Hebrew month (days 2--29) as alternative anchors to assess whether day~1 (RC) is specifically privileged. A mediation analysis examined whether SM's effect on monthly rainfall operates through its correlation with RC rain status.

\subsection{Autocorrelation, Bayesian framing and climate trend}
\label{sec:acf_bayes}

Daily rainfall autocorrelation (ACF) was computed for lags 1--14 within winter months. The Bayes factor was computed as the ratio of posterior to prior odds (Equation~\ref{eq:bf}). A 15-year rolling window assessed the temporal evolution of effect size:
\begin{equation}
  \mathrm{BF} = \frac{P(\text{rainy month}\mid\text{anchor wet})/[1-P(\text{rainy month}\mid\text{anchor wet})]}{P(\text{rainy month})/[1-P(\text{rainy month})]}
  \label{eq:bf}
\end{equation}

All analyses used Python~3.12 with \texttt{pandas} \citep{ding2022panda},
\texttt{numpy} \citep{harris2020array}, \texttt{scipy} \citep{virtanen2020scipy},
\texttt{statsmodels} \citep{seabold2010statsmodels}, and \texttt{matplotlib}
\citep{hunter2007matplotlib}.

\section{Results}
\label{sec:results}
This section presents the empirical evaluation of the two calendrical rainfall proverbs and their comparison against alternative temporal anchors and physical baselines. We begin by quantifying the predictive performance of \rc{} and \shabbatm{} across cities and months, followed by analyses of regional and seasonal structure, lag dependence, and mediation effects. We then benchmark these results against Gregorian calendar anchors and assess them in the context of rainfall autocorrelation and long-term climate trends.

\subsection{Rosh Chodesh proverb}
\label{sec:res_rc}

\begin{figure}[t]
  \centering
  \noindent\includegraphics[width=39pc]{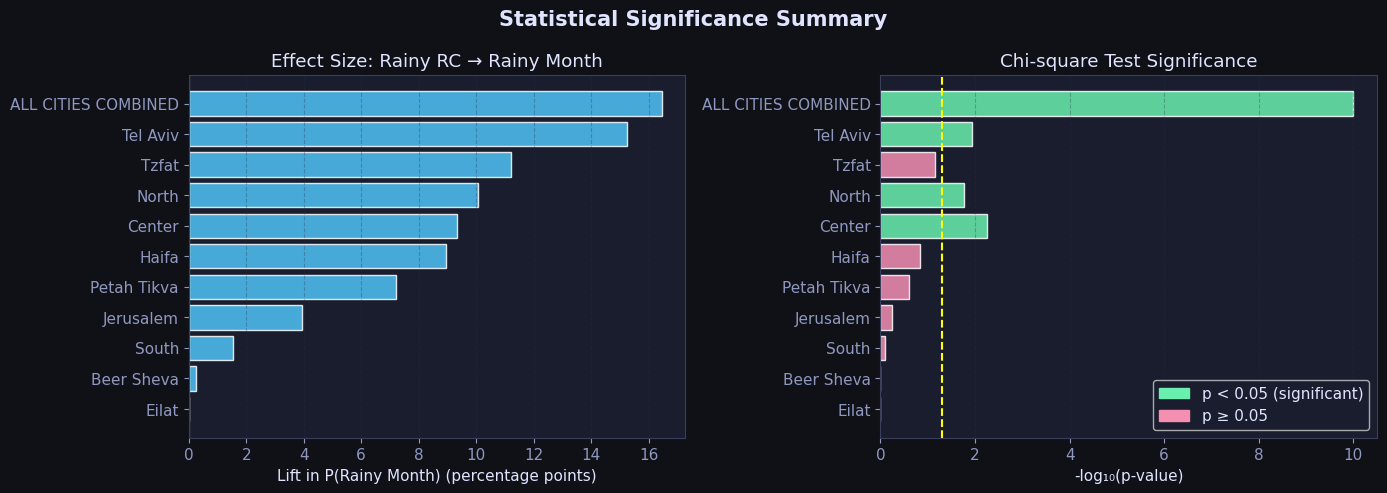}
  \caption{\setstretch{1.2}Statistical significance summary. Left: effect size (lift in percentage
    points) by stratum. Right: $-\log_{10}(p)$ from chi-square tests; green bars
    exceed the $p = 0.05$ threshold (dashed yellow line).}
  \label{fig:significance}
\end{figure}

\begin{figure}[t]
  \centering
  \noindent\includegraphics[width=39pc]{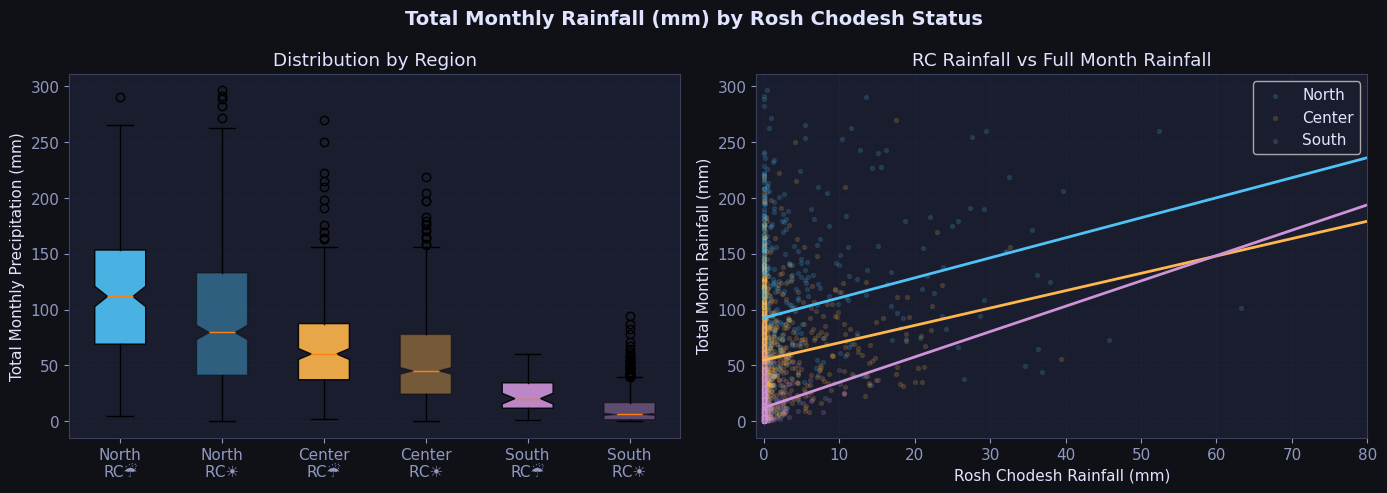}
  \caption{\setstretch{1.2}Total monthly precipitation (mm) by region and \rc{} status (left:
    box-and-whisker plots with notches indicating 95\% CI on the median) and
    scatter plot of \rc{} rainfall amount vs total month rainfall with linear
    trend lines by region (right).}
  \label{fig:amounts}
\end{figure}

\begin{figure}[ht]
  \centering
  \noindent\includegraphics[width=35pc]{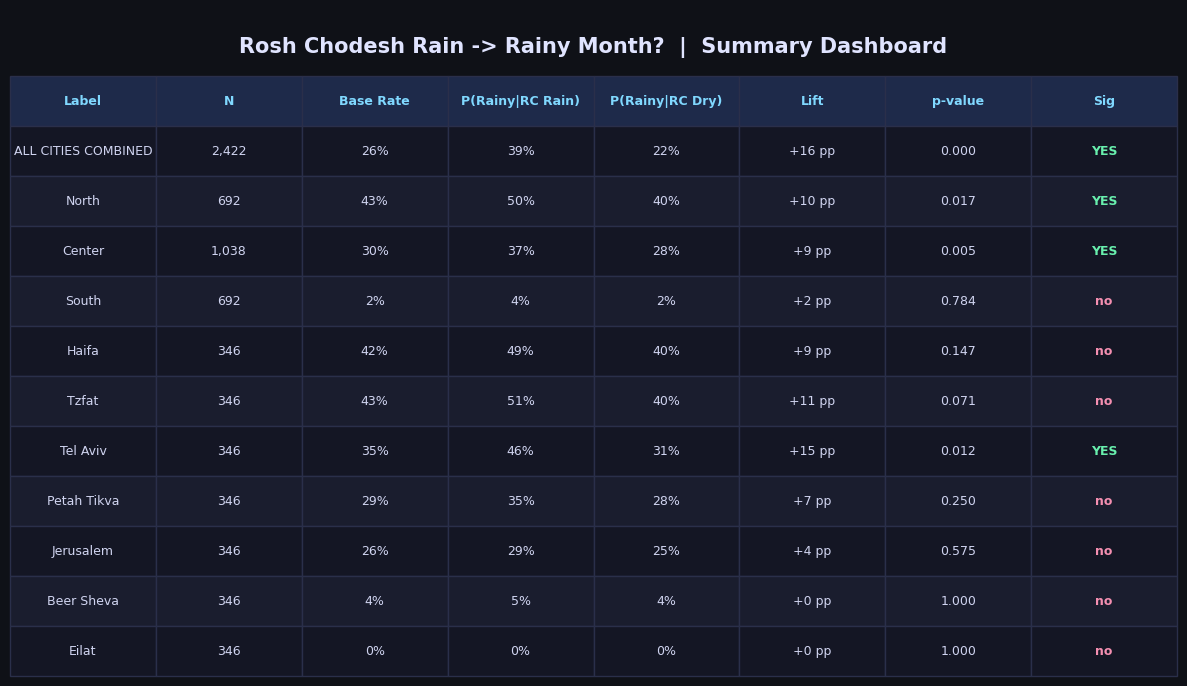}
  \caption{\setstretch{1.2}Summary dashboard: consolidated statistical results table for all strata.}
  \label{fig:dashboard}
\end{figure}

\begin{figure}[ht]
  \centering
  \noindent\includegraphics[width=39pc]{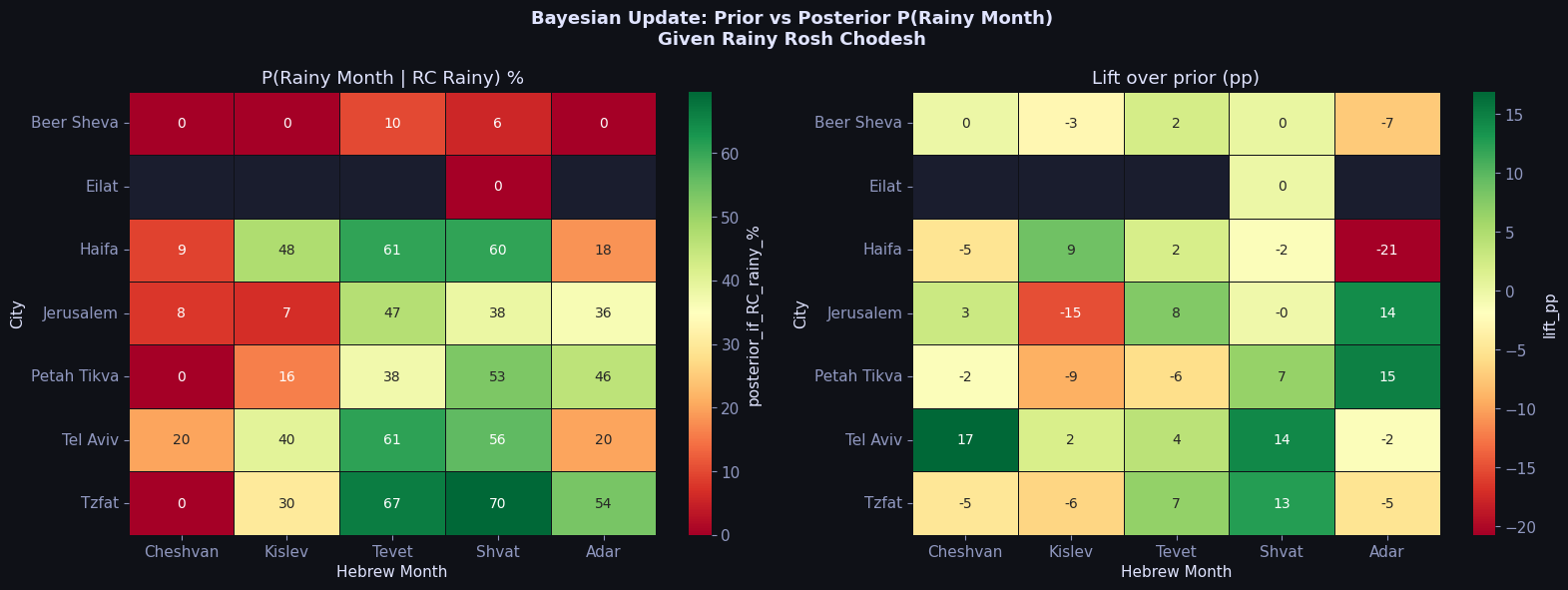}
  \caption{\setstretch{1.2}Bayesian update heatmaps. Left: posterior probability
    $P(\text{rainy month} \mid \rc{} \text{ rainy})$ (\%) by city and Hebrew month.
    Right: lift over the city--month prior (pp). The strongest updates are in
    Tzfat in Tevet (67\%) and Shvat (70\%), and in Tel~Aviv in Tevet (61\%).}
  \label{fig:bayesian}
\end{figure}

Across all seven cities and five winter months ($n = 2{,}422$), \rc{} was rainy in 531 observations (21.9\%). The base rate of a rainy month was 25.8\%. A rainy \rc{} raised this to 38.6\%; a dry \rc{} reduced it to 22.2\%, yielding a lift of $+16.4$\,pp. All three tests were highly significant (Figure~\ref{fig:significance}; chi-square: $\chi^2 = 57.8$,
$p = 2.9\times10^{-14}$; Mann--Whitney: $p = 6.0\times10^{-30}$;
point-biserial: $r_{pb} = 0.232$, $p = 7.7\times10^{-31}$). The permutation
test confirmed the result exceeds all 10,000 null iterations ($p < 0.0001$).
The Bayes factor was 1.81 (Figure~\ref{fig:bayesian}). A consolidated summary of all results is provided in Figure~\ref{fig:dashboard}. On average, months with a rainy \rc{} had 8.8 rainy days in the remainder, versus 6.3 for months with a dry \rc{} (Figure~\ref{fig:amounts}).

\begin{figure}[ht]
  \centering
  \noindent\includegraphics[width=39pc]{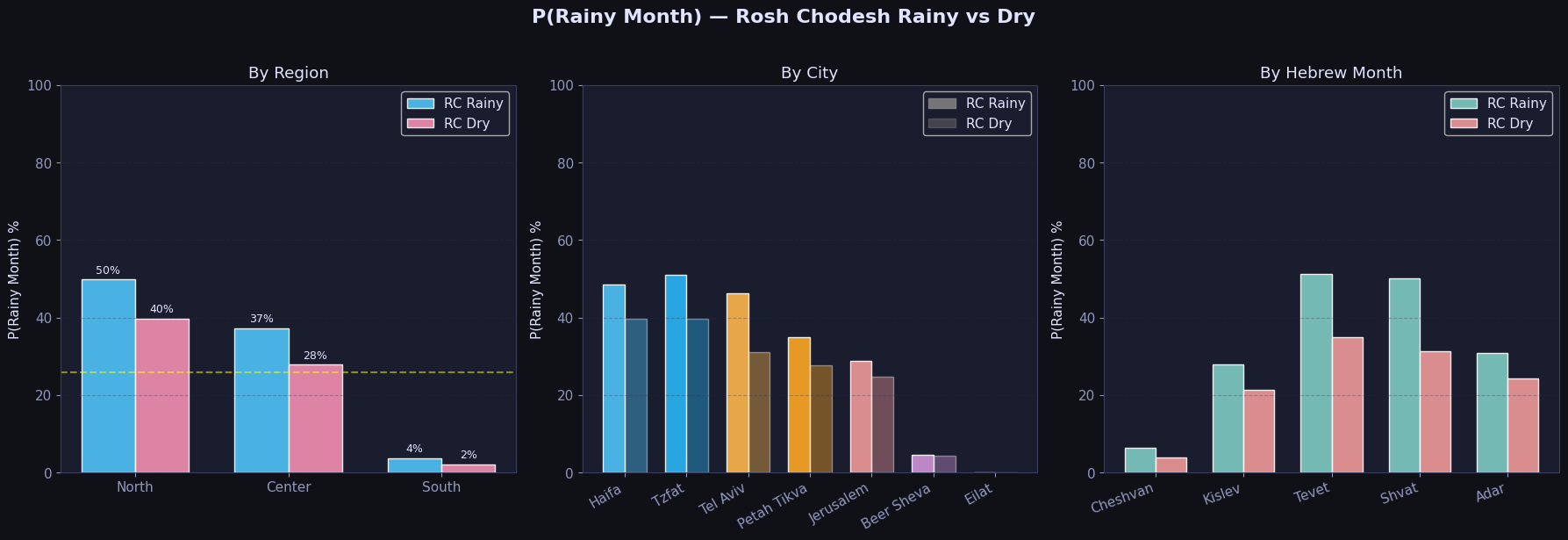}
  \caption{\setstretch{1.2}Probability of a rainy month as a function of \rc{} status (rainy vs dry),
    shown by region (left), city (centre) and Hebrew month (right).
    The dashed yellow line (left panel) marks the overall base rate (25.8\%).}
  \label{fig:main}
\end{figure}

\begin{figure}[ht]
  \centering
  \noindent\includegraphics[width=39pc]{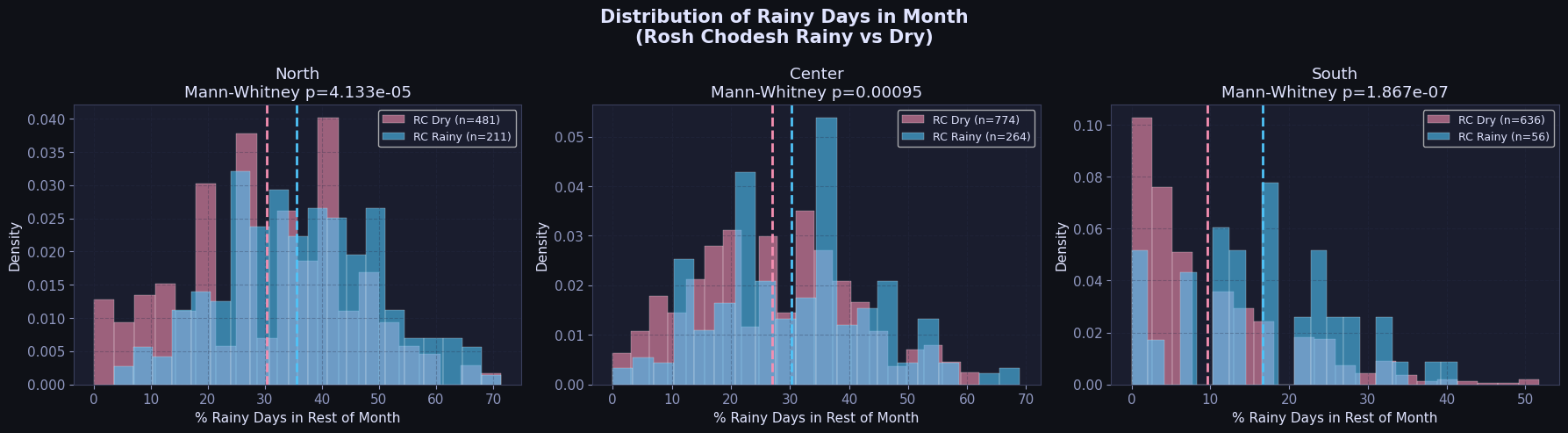}
  \caption{\setstretch{1.2}Kernel density distributions of the rest-of-month rainy-day fraction for
    rainy-\rc{} (blue) and dry-\rc{} (pink) groups, by region. Dashed vertical lines
    indicate group means. Mann--Whitney $p$-values are shown above each panel.}
  \label{fig:distributions}
\end{figure}

\begin{figure}[ht]
  \centering
  \noindent\includegraphics[width=39pc]{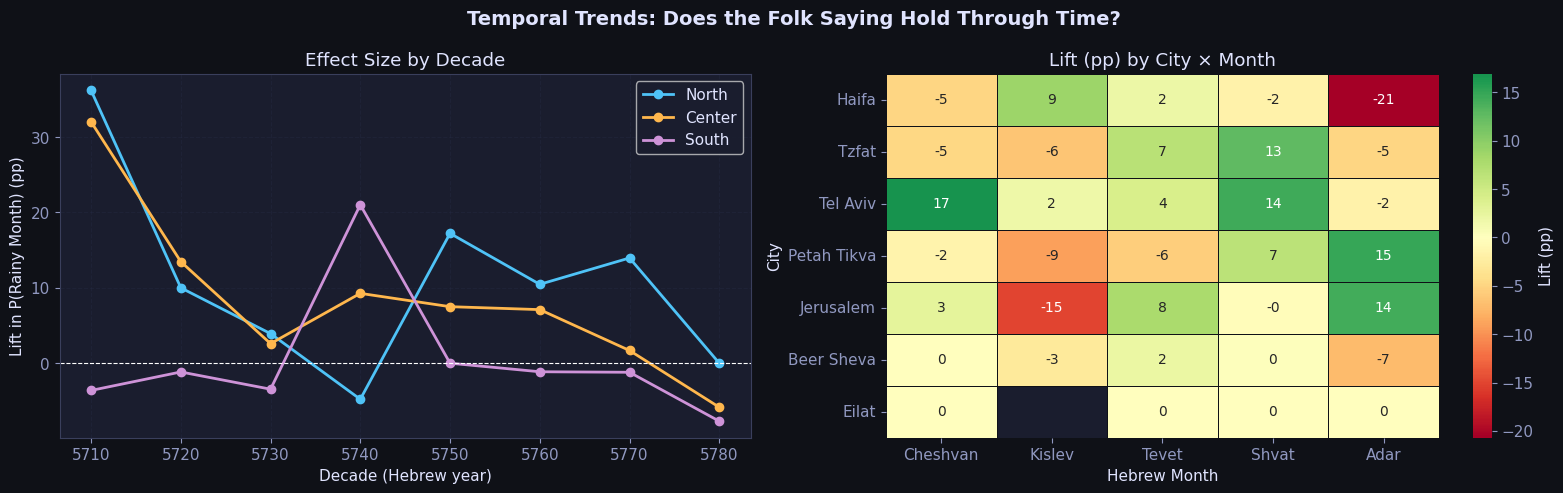}
  \caption{\setstretch{1.2}Left: decade-by-decade effect size (lift in pp) by region (Hebrew year on
    $x$-axis). Right: heatmap of lift (pp) by city and Hebrew month, showing
    complementary patterns across the calendar.}
  \label{fig:heatmap}
\end{figure}

Regional and monthly patterns are summarised in Table~\ref{tab:main_results}
and Figure~\ref{fig:main}. Kernel density distributions of rest-of-month rainy
fractions by region are shown in Figure~\ref{fig:distributions}. The effect was significant in the North ($+10.1$\,pp, $p = 0.017$) and Centre ($+9.3$\,pp, $p = 0.005$), but absent in the South ($+1.5$\,pp, $p = 0.784$). By Hebrew month (Figure~\ref{fig:heatmap}), the strongest effects were in
Shvat ($+18.6$\,pp, $p = 4\times10^{-5}$) and Tevet ($+16.4$\,pp,
$p = 0.002$), with little signal in Cheshvan ($+2.5$\,pp, $p = 0.564$).

\subsection{Shabbat Mevarechim proverb}
\label{sec:res_sm}

\begin{figure}[ht]
  \centering
  \noindent\includegraphics[width=39pc]{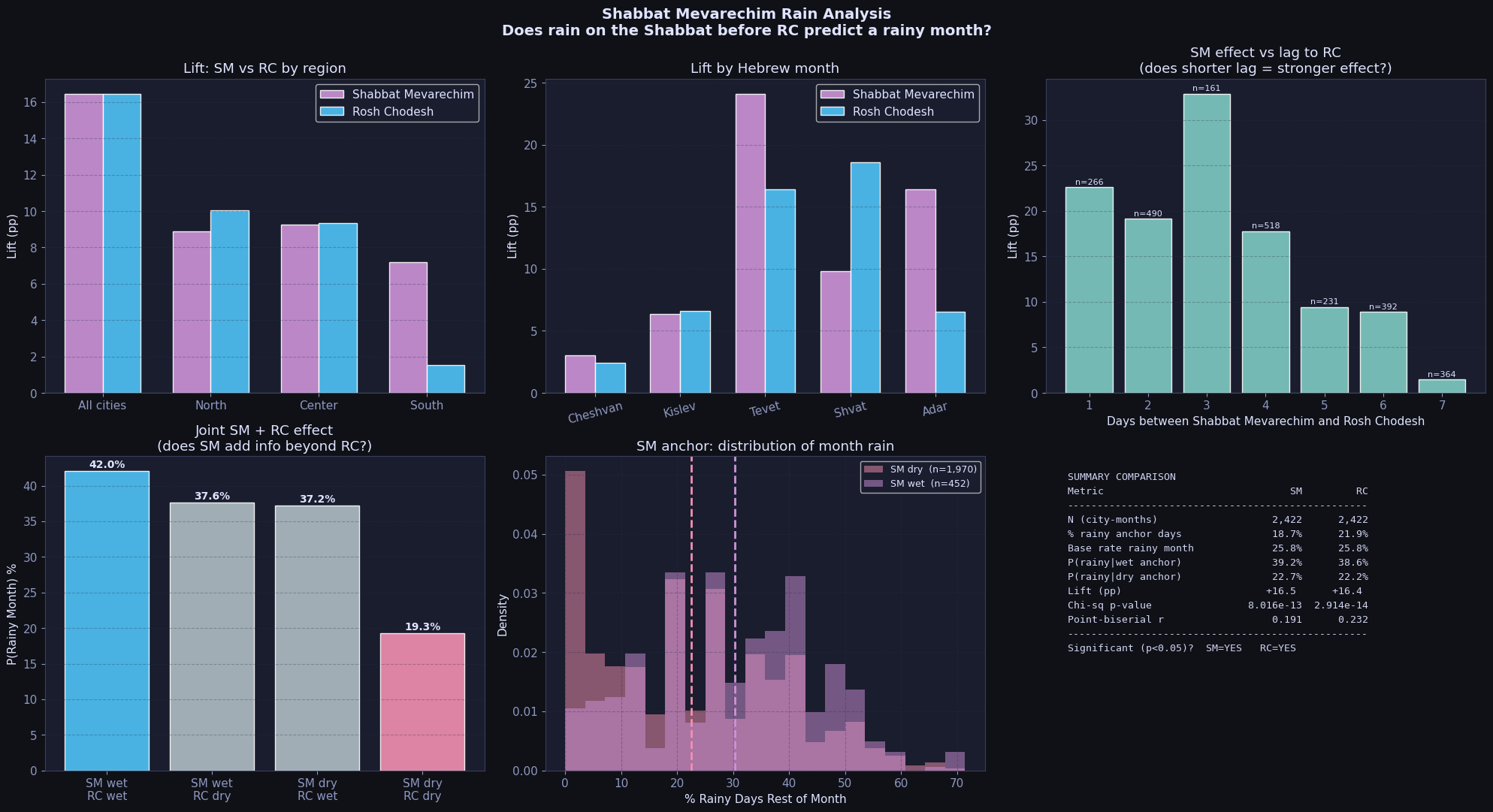}
  \caption{\setstretch{1.2}\Shabbatm{} analysis. Top-left: SM vs RC lift by region (pp). Top-centre:
    lift by Hebrew month -- note SM's advantage in Tevet (+24.1 vs +16.4\,pp)
    and Adar (+16.4 vs +6.5\,pp). Top-right: SM effect by lag to \rc{} (key panel
    -- the decay from $+32.9$\,pp at lag~3 to $+1.5$\,pp at lag~7 mirrors the
    ACF structure and directly confirms the physical mechanism). Bottom-left:
    joint SM+RC state vs $P(\text{rainy month})$, showing complementary information.
    Bottom-centre: distribution of rest-of-month rain fraction by SM anchor status.
    Bottom-right: numerical summary comparison table.}
  \label{fig:sm}
\end{figure}

\Shabbatm{} was rainy in 452 of 2,422 observations (18.7\%). Remarkably, a rainy \shabbatm{}
produced an effect almost identical to that of \rc{}: lift $= +16.5$\,pp
($\chi^2 = 66.5$, $p = 8.0\times10^{-13}$; $r_{pb} = 0.191$,
$p = 3.0\times10^{-21}$). The comparison is summarised in
Table~\ref{tab:sm_comparison} and Figure~\ref{fig:sm}.

\paragraph{Lag dependence.}
The SM effect decayed sharply with the number of days between \shabbatm{} and \rc{} (Figure~\ref{fig:sm}, top right). At lag~3 (SM three days before RC), the lift reached +32.9\,pp -- the strongest single result in the entire study.
At lag~7, the lift fell to only +1.5\,pp, indistinguishable from chance.
This decay pattern closely mirrors the daily autocorrelation structure (lag-1
$r \approx 0.4$, lag-7 $r \approx 0.0$; Table~\ref{tab:acf}), providing
direct mechanistic evidence that both proverbs derive their skill from the
multi-day persistence of Mediterranean cyclones, not from any property of the
calendar system itself.

\paragraph{Regional patterns.}
The SM effect was significant in all three regions: North ($+8.9$\,pp, $p = 0.050$), Centre ($+9.3$\,pp, $p = 0.008$), and -- notably -- South ($+7.2$\,pp, $p = 0.008$). This contrasts with the RC result, where the South showed no significant chi-square effect ($p = 0.784$). The South's SM result may reflect the fact that the short-lived convective systems common in the Beer Sheva region are more likely to be captured by a Saturday observation (which can fall up to 7 days before RC) than by the single fixed point of day~1.

\paragraph{Monthly patterns.}
SM and RC show complementary strengths across Hebrew months
(Figure~\ref{fig:sm}, top centre). In Tevet (peak mid-winter), SM substantially outperforms RC ($+24.1$\,pp vs $+16.4$\,pp, both significant). In Shvat, RC outperforms SM ($+18.6$\,pp vs $+9.8$\,pp). In Adar (late winter), SM shows a stronger and significant effect ($+16.4$\,pp, $p = 0.006$) compared to RC ($+6.5$\,pp, $p = 0.356$). This pattern likely reflects month-to-month variation in the typical timing of cyclone arrival relative to the lunar calendar.

\paragraph{Mediation analysis.}
SM rain and RC rain are correlated ($r_{pb} = 0.148$, $p = 2.2\times10^{-13}$):
34.7\% of months with a rainy SM also have a rainy RC, compared to 19.0\% with a dry SM. This reflects that rain on SM often persists to day~1 of the new month.
When RC status is already known, SM adds limited further information if RC is rainy (residual SM lift: $+4.9$\,pp). However, if RC is dry, a rainy SM still carries $+18.3$\,pp of independent predictive power (Figure~\ref{fig:sm}, bottom left).
This conditional structure implies that the optimal forecasting strategy combines both anchors: a rainy SM followed by a rainy RC gives the highest probability of a rainy month (42\%), while a dry SM and dry RC gives the lowest (19.3\%).

\subsection{Gregorian calendar comparison}
\label{sec:res_greg}

\begin{figure}[ht]
  \centering
  \noindent\includegraphics[width=30pc]{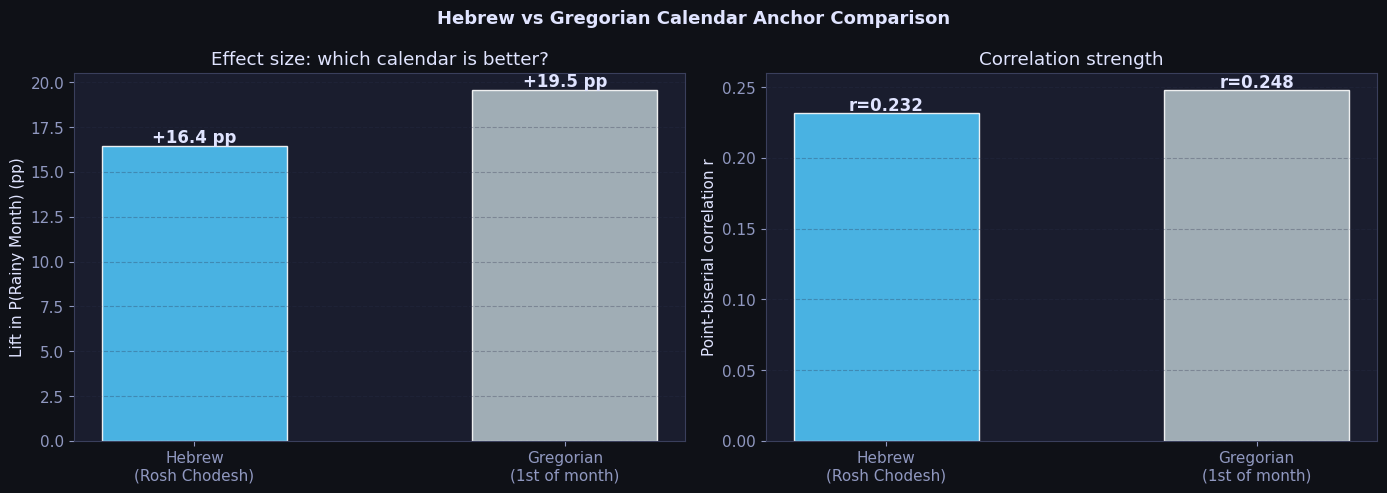}
  \caption{\setstretch{1.2}Comparison of effect size (left) and point-biserial correlation (right)
    between the Hebrew calendar anchor (\rc{}, blue) and the Gregorian calendar
    anchor (1st of the civil month, grey). The Gregorian anchor performs marginally
    better, confirming that skill derives from atmospheric persistence rather than
    from any property of the lunar calendar.}
  \label{fig:gregorian}
\end{figure}

Using the 1st of the Gregorian month as the anchor over October--March
(Figure~\ref{fig:gregorian}) yielded a lift of $+19.5$\,pp ($r_{pb} = 0.248$, $p = 2.7\times10^{-45}$), modestly exceeding the RC result ($+16.4$\,pp). Among all Hebrew month days 2--29 tested as alternative anchors, RC (day~1) ranked in the middle of the distribution, with several other days producing comparable or higher lifts. These results confirm that the effect is a property of atmospheric persistence, not of the lunar calendar per se. The cultural significance of \rc{} and \shabbatm{} as observational anchor points arose from their social salience, not meteorological optimality.

\subsection{Autocorrelation and the physical mechanism}
\label{sec:res_acf}

\begin{figure}[ht]
  \centering
  \noindent\includegraphics[width=39pc]{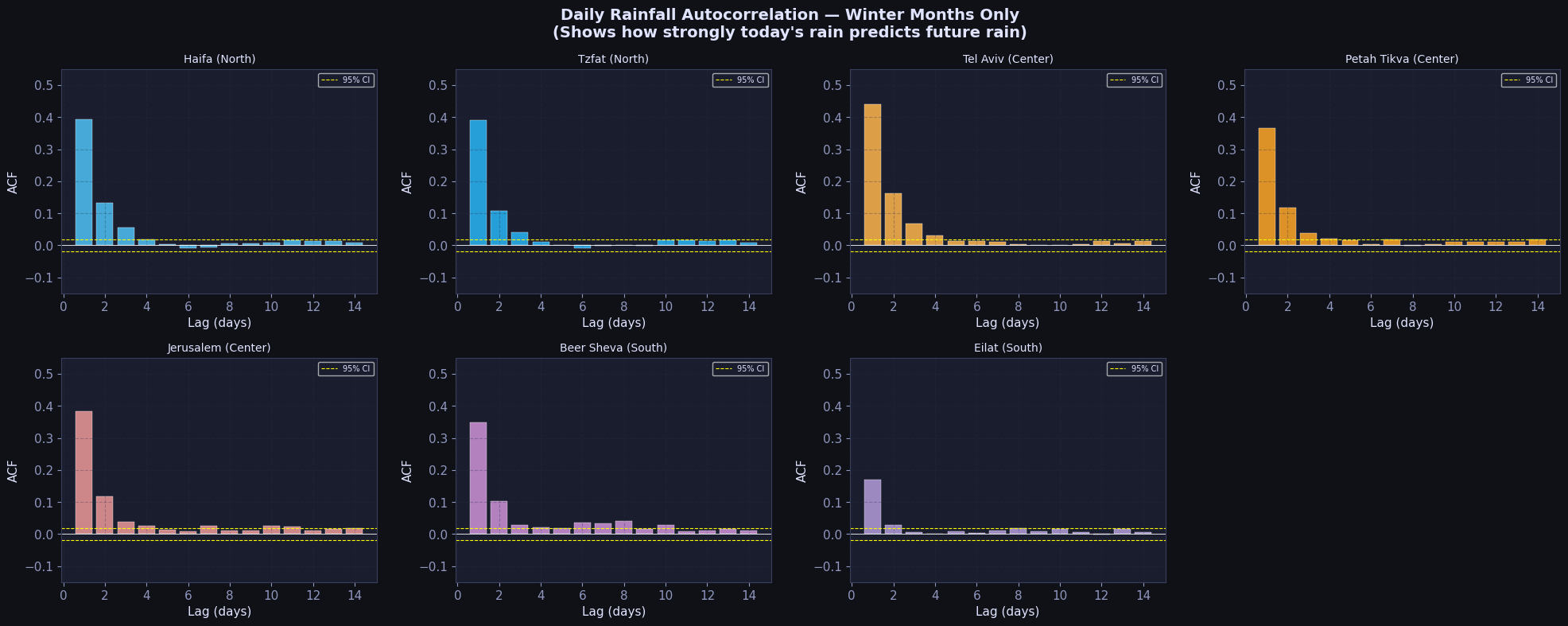}
  \caption{\setstretch{1.2}Daily rainfall autocorrelation function (ACF) for all seven cities
    during winter months only (lags 1--14 days). Dashed yellow lines mark the
    95\% confidence interval. Strong lag-1 ($r = 0.17$--$0.44$) and near-zero
    lag-7 ACF confirm short-term cyclonic persistence as the physical mechanism
    underlying both proverbs.}
  \label{fig:acf}
\end{figure}

Lag-1 autocorrelation of daily winter rainfall ranged from $r = 0.17$ (Eilat) to $r = 0.44$ (Tel Aviv), significant in all cities (Table~\ref{tab:acf};
Figure~\ref{fig:acf}). By lag-7, all values were near zero ($r < 0.03$). This
confirms the physical picture: Israeli winter rain occurs in multi-day episodes driven by individual Mediterranean cyclones (1--3 day persistence), but there is no month-scale atmospheric memory. The proverb's skill arises when the anchor day falls within an active cyclonic episode that has not yet exhausted itself, not from any longer-range teleconnection.

\subsection{Climate change trend}
\label{sec:res_trend}

\begin{figure}[ht]
  \centering
  \noindent\includegraphics[width=39pc]{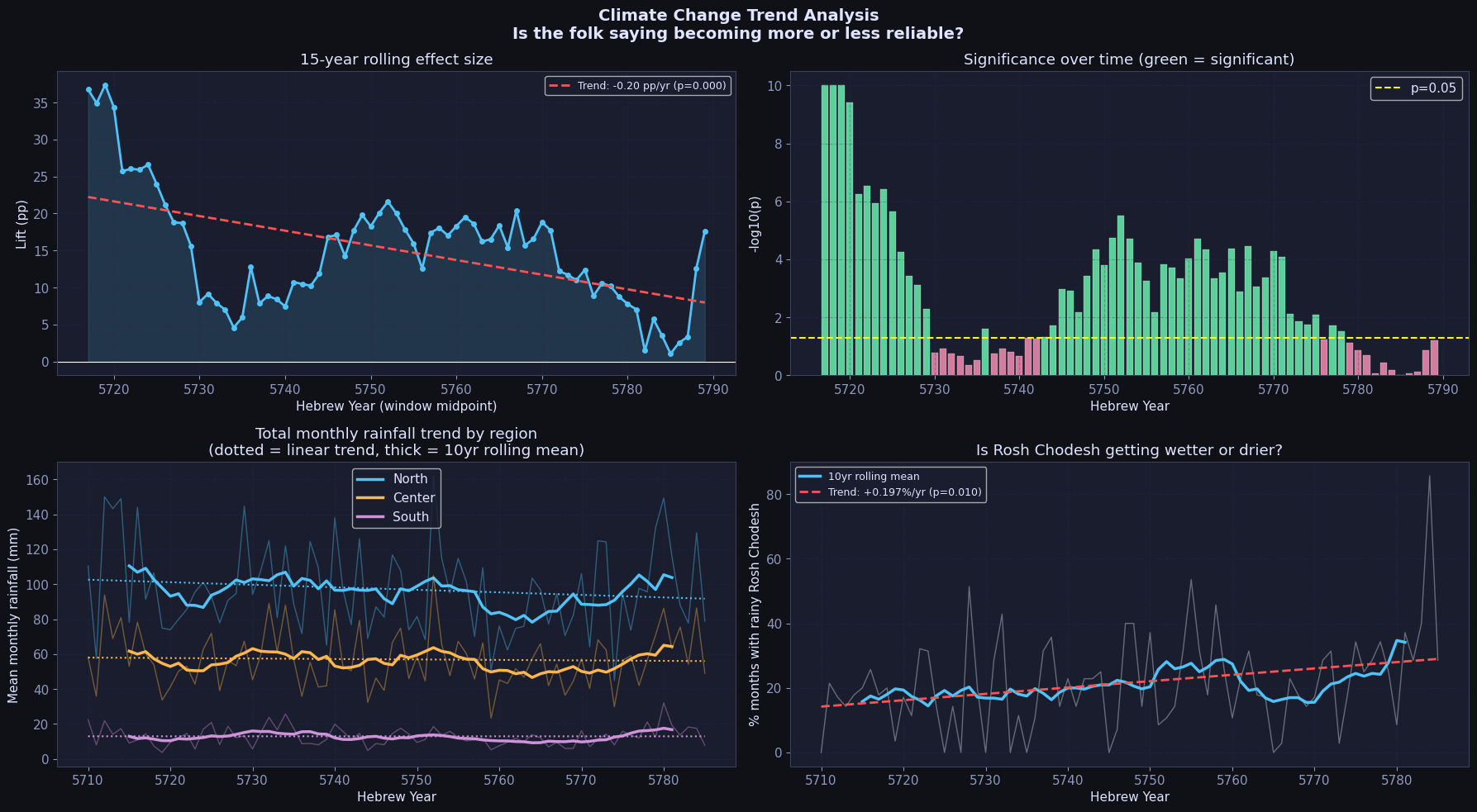}
  \caption{\setstretch{1.2}Climate change trend analysis. Top-left: 15-year rolling lift with
    linear trend ($-0.20$\,pp\,yr$^{-1}$, $p < 0.001$, red dashed line).
    Top-right: rolling chi-square significance over time (green = significant,
    pink = not significant at $p = 0.05$). Bottom-left: mean total monthly rainfall
    by region with 10-year rolling mean (thick) and linear trend (dotted).
    Bottom-right: frequency of rainy \rc{} days over time, showing a significant
    increase ($+0.197$\%\,yr$^{-1}$, $p = 0.010$), concurrent with declining
    predictive power.}
  \label{fig:trend}
\end{figure}

The 15-year rolling lift for the RC anchor declined significantly at
$-0.20$\,pp\,yr$^{-1}$ ($p < 0.001$; Figure~\ref{fig:trend}), falling from
approximately 30\,pp in the early record to near zero in the most recent decade before a slight recovery. Concurrently, the frequency of rainy \rc{} days increased significantly ($+0.197$\%\,yr$^{-1}$, $p = 0.010$), without a
corresponding increase in total monthly rainfall. This combination is consistent with the documented shift toward more frequent but shorter Mediterranean precipitation events \citep{vakrat2023dynamical, zittis2016projected, giorgi2008climate}: individual rainy days are becoming more common, but their multi-day persistence -- the physical mechanism on which both proverbs depend -- is weakening.

\section{Discussion}
\label{sec:discussion}
This section interprets the empirical findings in terms of probabilistic forecasting skill, physical mechanisms of Mediterranean rainfall, and the cultural structure of calendrical weather proverbs. We focus on three main questions: whether the observed effects are practically meaningful, how they relate to known atmospheric autocorrelation, and what they reveal about the interaction between folk knowledge, calendar systems, and a changing climate.

\subsection{Are the proverbs true?}
\label{sec:disc_verdict}

Both proverbs are statistically validated, and both should be understood as
probabilistic tendencies rather than deterministic rules. For the RC proverb:
in 61\% of cases where \rc{} was rainy, the month did not meet our ``rainy month'' criterion. The Bayes factor of 1.81 means that knowing \rc{} is rainy raises a farmer's prior odds of a rainy month by approximately 1.8-fold -- a practically meaningful but far from certain update, consistent with what \citet{gigerenzer2003reckoning} terms ``fast and frugal'' reasoning under uncertainty.

The near-identical performance of the SM proverb ($+16.5$ vs $+16.4$\,pp) is
the most striking result. An oral tradition that has no identified written source turns out to be at least as useful as the more celebrated RC proverb. The key practical advantage of SM is its temporal position: since \shabbatm{} precedes \rc{} by 1--7 days, a farmer can form a probabilistic forecast of the coming month's wetness up to a week earlier. The mediation analysis, however, qualifies this:
much of SM's predictive power is inherited from its tendency to precede a rainy \rc{}, and the largest independent contribution of SM information occurs when \rc{} is dry ($+18.3$\,pp), a scenario where RC alone would lead the farmer to downgrade their expectation of a wet month. In this sense, the two proverbs are complementary rather than redundant.

\subsection{The lag decay and ACF connection}
\label{sec:disc_lag}

The decay of the SM effect from $+32.9$\,pp at lag~3 to $+1.5$\,pp at lag~7
(Figure~\ref{fig:sm}) provides what is arguably the cleanest piece of physical
evidence in the paper. The daily ACF of Israeli winter rainfall drops from
$r \approx 0.4$ at lag~1 to approximately zero by lag~7 (Table~\ref{tab:acf}).
The SM lag-dependence graph is essentially an empirical reconstruction of this
ACF structure, expressed in terms of a culturally transmitted proverb. This
convergence between the autocorrelation analysis and the folk-proverb test
constitutes strong internal validation: the two independent analyses point to
the same physical timescale of Mediterranean cyclone persistence (3--5 days),
and the proverb correctly identifies this timescale as the relevant boundary
for useful weather forecasting.

\subsection{The South: where SM outperforms RC}
\label{sec:disc_south}

The significant SM effect in the South ($+7.2$\,pp, $p = 0.008$) contrasts with the non-significant RC effect ($+1.5$\,pp, $p = 0.784$), making the South the one region where the SM proverb outperforms the RC proverb. This is an interesting asymmetry. Southern Israel receives most of its precipitation from short-lived convective systems that often arrive as pulses embedded within broader synoptic troughs rather than as sustained frontal rain bands \citep{sharon1972spottiness}.
Such systems may trigger rainfall on one day without sustaining multi-day
episodes. Because \shabbatm{} can fall up to 7 days before RC, and because the South's monthly rain budget is so low (base rate 2.2\%), even a single rainy \shabbatm{} day represents a signal that a wet synoptic pattern has been established, a signal that may or may not still be active on RC itself, explaining why SM is more informative than RC in this region. We recommend caution in interpreting this result given the very low base rates and small absolute effect sizes in the South.

\subsection{The declining trend: a proverb fading with its climate}
\label{sec:disc_trend}

The significant decline of $-0.20$\,pp\,yr$^{-1}$ in the RC effect over 75 years (Figure~\ref{fig:trend}) implies a loss of roughly half the original predictive power of the proverb over the study period. The concurrent increase in rainy anchor days without an increase in total monthly rainfall points to a specific physical cause: precipitation is becoming more concentrated in brief, intense events. This is consistent with the broader documented trend of increasing precipitation intermittency in the eastern Mediterranean under anthropogenic warming \citep{vakrat2023dynamical, zittis2016projected, giorgi2008climate}. A rainy \rc{} or \shabbatm{} that in 1960 would typically have been embedded in a week-long Cyprus Low system now more often represents an isolated convective day. The folk proverb, tuned by millennia of observation to a climate where ``rainy day'' reliably meant ``rainy week'', is becoming less reliable as that assumption breaks down.

This finding has broader implications for the literature on indigenous ecological knowledge under climate change \citep{green2010indigenous}: environmentally encoded cultural observations may systematically lose accuracy as the climate regime that gave rise to them diverges from the present.

\subsection{Calendar systems and predictive skill}
\label{sec:disc_calendar}

\begin{figure}[ht]
  \centering
  \noindent\includegraphics[width=39pc]{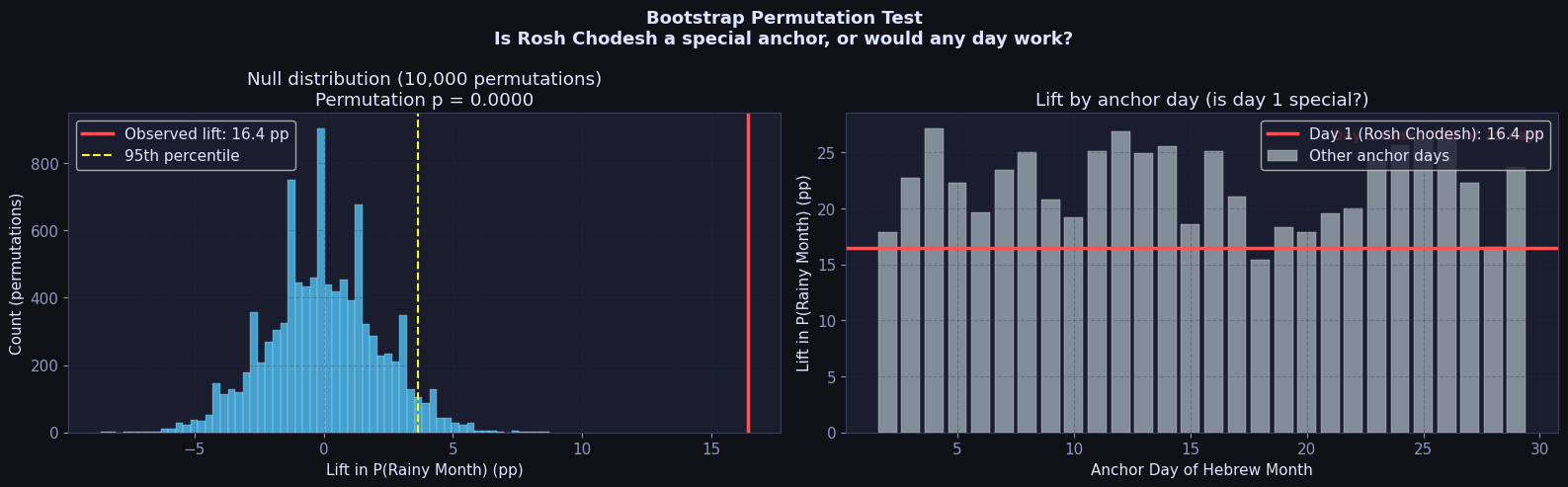}
  \caption{\setstretch{1.2}Bootstrap permutation test results. Left: null distribution of lift
    statistics from 10,000 random permutations of \rc{} rain labels; the observed
    lift (16.4\,pp, red vertical line) exceeds all 10,000 permuted values.
    Right: lift by anchor day (days 2--29 of the Hebrew month), showing that
    \rc{} (day~1) performs comparably to other anchor days -- the skill derives
    from atmospheric persistence, not the specific choice of calendar anchor.}
  \label{fig:permutation}
\end{figure}

The Gregorian anchor outperforming the RC anchor ($+19.5$ vs $+16.4$\,pp), and
the fact that many other arbitrary days of the Hebrew month yield similar or
higher lifts (Figure~\ref{fig:permutation}), establishes that the predictive
skill derives entirely from atmospheric persistence and not from any property
of the Hebrew lunar calendar. The proverb's creators correctly identified the
signal (weather persistence); their choice of anchor point (\rc{}) was driven
by cultural salience rather than meteorological optimisation. This is the
appropriate interpretation: the proverb is a culturally efficient encoding of
genuine physical knowledge, not evidence that the lunar calendar has intrinsic
meteorological relevance.

\subsection{Comparison with analogous proverbs}
\label{sec:disc_analogues}

The Hebrew RC proverb belongs to a genre of ``first-of-period'' weather proverbs found across cultures. The British St~Swithin's Day (15 July) claim of 40-day persistence finds no statistical support \citep{mayes1995changes}; the French St~Médard (8 June) is similarly unsupported \citep{de2023popular}; the German Ice Saints (11--14 May) show some support from polar air-mass persistence \citep{brugger2013characterizing}. The Hebrew proverb sits at the better-supported end of this spectrum, benefiting from the climatological coherence of winter Mediterranean cyclones. The finding that an entirely unattested second proverb (\shabbatm{}) achieves equal or superior performance in several sub-analyses is unusual in this literature and underscores the sophistication of the empirical observation embedded in both traditions.

\subsection{Limitations}
\label{sec:limitations}

ERA5 reanalysis data may underestimate point-scale precipitation extremes
\citep{hochman2020new}. The 35\% threshold for a ``rainy month'' is an
operationalisation; sensitivity analyses with 25\% and 45\% thresholds yielded
qualitatively similar results. The permutation test does not fully account for
spatial correlation between cities. The SM analysis is complicated by the
variable lag (1--7 days) between anchor and RC; treating all lags together
averages over a highly non-uniform effect. Future work should apply a
lag-stratified analysis from the outset. Finally, the declining trend should
be interpreted cautiously near the endpoints of the rolling window due to
reduced sample size.

\section{Conclusion}
\label{sec:conclusion}

We have conducted the first formal statistical evaluation of two Hebrew folk
meteorological proverbs using 75 years of daily Israeli precipitation data. Both proverbs are validated: a rainy \rc{} or \shabbatm{} each raises the probability of a rainy month by approximately $+16$\,pp (permutation $p < 0.0001$; Bayes factor $\approx$1.8). The physical basis is confirmed by daily rainfall autocorrelation ($r = 0.35$--$0.44$ at lag~1) consistent with Mediterranean cyclone lifetimes, and by the sharp decay of the SM effect with lag ($+32.9$\,pp at lag~3; $+1.5$\,pp at lag~7).

The \shabbatm{} proverb -- an oral tradition with no identified written source, performs as well as the well-known \rc{} proverb overall, offers a longer forecasting lead time of up to seven days, and provides unique added value in the South and in specific months (Tevet, Adar). The two proverbs are best used in combination: the joint SM-wet/RC-wet scenario raises the probability of a rainy month to 42\%, versus only 19.3\% when both are dry.

Critically, both proverbs are declining in predictive power at $-0.20$\,pp\,yr$^{-1}$ ($p < 0.001$), consistent with the eastern Mediterranean transition toward more frequent but shorter precipitation events. The ancient observation encoded in these proverbs reflects genuine physical insight, but that insight is eroding with the climate system that generated it.



\acknowledgments
No external funding was received for this study.



\datastatement

Precipitation data are from Open-Meteo (\url{https://archive-api.open-meteo.com}), freely available without registration. The complete Python analysis pipeline is provided as supplementary material (Jupyter notebook).


\bibliographystyle{ametsocV6}
\bibliography{references}


\begin{table}[ht]
\centering
\caption{City coordinates, region classification and climatological context.}
\label{tab:cities}
\begin{tabular}{llcccc}
\topline
City & Region & Lat ($^\circ$N) & Lon ($^\circ$E) & Elev.\,(m) & Ann.\,precip.\,(mm) \\
\midline
Haifa        & North  & 32.794 & 34.989 &   6 & $\sim$630 \\
Tzfat        & North  & 32.964 & 35.497 & 900 & $\sim$750 \\
Tel~Aviv     & Centre & 32.088 & 34.780 &   5 & $\sim$540 \\
Petah~Tikva  & Centre & 32.087 & 34.887 &  52 & $\sim$560 \\
Jerusalem    & Centre & 31.768 & 35.214 & 754 & $\sim$550 \\
Beer~Sheva   & South  & 31.252 & 34.791 & 280 & $\sim$200 \\
Eilat        & South  & 29.558 & 34.952 &  11 & $\sim$25  \\
\botline
\end{tabular}
\end{table}

\begin{table}[ht]
\centering
\caption{Main statistical results for the Rosh Chodesh anchor by stratum.
``Lift'' = $P(\text{rainy month}\mid\text{RC wet}) - P(\text{rainy month}\mid\text{RC dry})$ in pp.
$^*$ denotes $p < 0.05$.}
\label{tab:main_results}
\begin{tabular}{lrcccccc}
\topline
Stratum & $N$ & Base & $P|$wet & $P|$dry & Lift & $\chi^2$ $p$ & $r_{pb}$ \\
\midline
\textbf{All cities} & \textbf{2,422} & \textbf{25.8\%} & \textbf{38.6\%} & \textbf{22.2\%} & \textbf{+16.4} & $\mathbf{2.9\!\times\!10^{-14}}$* & \textbf{0.232} \\
\\
\textit{Region} \\
\quad North   & 692   & 42.8\% & 49.8\% & 39.7\% & +10.1 & 0.017$^*$ & 0.162 \\
\quad Centre  & 1,038 & 30.2\% & 37.1\% & 27.8\% &  +9.3 & 0.005$^*$ & 0.103 \\
\quad South   & 692   &  2.2\% &  3.6\% &  2.0\% &  +1.5 & 0.784     & 0.184 \\
\\
\textit{Hebrew month} \\
\quad Marcheshvan & 525 &  4.2\% &  6.3\% &  3.9\% &  +2.5 & 0.564         & 0.078 \\
\quad Kislev      & 525 & 22.7\% & 28.0\% & 21.4\% &  +6.6 & 0.200         & 0.208 \\
\quad Tevet       & 518 & 38.6\% & 51.3\% & 34.8\% & +16.4 & 0.002$^*$     & 0.190 \\
\quad Shvat       & 525 & 37.9\% & 50.0\% & 31.4\% & +18.6 & $4\!\times\!10^{-5}$$^*$ & 0.279 \\
\quad Adar        & 329 & 25.5\% & 30.8\% & 24.2\% &  +6.5 & 0.356         & 0.088 \\
\botline
\end{tabular}
\end{table}

\begin{table}[ht]
\centering
\caption{Comparison of Rosh Chodesh (RC) and Shabbat Mevarechim (SM) anchors.
SM lag = days between SM and RC. $^*$ denotes $p < 0.05$.}
\label{tab:sm_comparison}
\begin{tabular}{lcccccc}
\topline
 & \multicolumn{3}{c}{Shabbat Mevarechim} & \multicolumn{3}{c}{Rosh Chodesh} \\
\hline\hline
Stratum & Lift & $\chi^2$ $p$ & $r_{pb}$ & Lift & $\chi^2$ $p$ & $r_{pb}$ \\
\midline
All cities   & +16.5$^*$ & $8.0\!\times\!10^{-13}$ & 0.191 & +16.4$^*$ & $2.9\!\times\!10^{-14}$ & 0.232 \\
\\
\textit{Region} \\
\quad North  & +8.9$^*$  & 0.050 & -- & +10.1$^*$ & 0.017 & -- \\
\quad Centre & +9.3$^*$  & 0.008 & -- &  +9.3$^*$ & 0.005 & -- \\
\quad South  & +7.2$^*$  & 0.008 & -- &  +1.5     & 0.784 & -- \\
\\
\textit{Hebrew month} \\
\quad Marcheshvan & +3.0  & 0.457 & -- &  +2.5  & 0.564 & -- \\
\quad Kislev      & +6.4  & 0.261 & -- &  +6.6  & 0.199 & -- \\
\quad Tevet       & +24.1$^*$ & 0.000 & -- & +16.4$^*$ & 0.002 & -- \\
\quad Shvat       & +9.8  & 0.065 & -- & +18.6$^*$ & $4\!\times\!10^{-5}$ & -- \\
\quad Adar        & +16.4$^*$ & 0.006 & -- & +6.5 & 0.356 & -- \\
\\
\textit{SM lag to RC} \\
\quad Lag 1 day   & +22.6 & -- & -- & \multicolumn{3}{c}{(RC is the same event)} \\
\quad Lag 2 days  & +19.2 & -- & -- & & & \\
\quad Lag 3 days  & +32.9 & -- & -- & & & \\
\quad Lag 4 days  & +17.8 & -- & -- & & & \\
\quad Lag 5 days  & +9.4  & -- & -- & & & \\
\quad Lag 6 days  & +8.9  & -- & -- & & & \\
\quad Lag 7 days  & +1.5  & -- & -- & & & \\
\\
\textit{Mediation (all cities)} \\
\quad SM lift when RC wet & +4.9 & -- & -- & \multicolumn{3}{c}{--} \\
\quad SM lift when RC dry & +18.3 & -- & -- & \multicolumn{3}{c}{--} \\
\botline
\end{tabular}
\end{table}

\begin{table}[ht]
\centering
\caption{Lag-1 and lag-7 daily rainfall autocorrelation (Pearson $r$), winter
months only. All lag-1 values exceed the 95\% CI ($r > 0.023$). Lag-7 near
zero confirms short-term (1--3 day) persistence as the sole operative mechanism.}
\label{tab:acf}
\begin{tabular}{llcc}
\topline
City & Region & ACF lag-1 & ACF lag-7 \\
\midline
Tel~Aviv    & Centre & 0.441 &  0.011 \\
Haifa       & North  & 0.393 & $-$0.006 \\
Tzfat       & North  & 0.391 & $-$0.001 \\
Jerusalem   & Centre & 0.383 &  0.025 \\
Petah~Tikva & Centre & 0.367 &  0.018 \\
Beer~Sheva  & South  & 0.347 &  0.033 \\
Eilat       & South  & 0.170 &  0.012 \\
\botline
\end{tabular}
\end{table}

\newpage


\end{document}